\documentclass{aa}  
\usepackage{graphicx}
\usepackage{txfonts}
\usepackage{amsmath}

\def\teff{${\rm T_{\rm eff}}$}

\def\bprp{$(\rm BP-RP)_{0}$}
\def\bpg{$(\rm BP-G)_{0}$}
\def\grp{$(\rm G-RP)_{0}$}
\def\bpk{$(\rm BP-K_{s})_{0}$}
\def\rpk{$(\rm RP-K_{s})_{0}$}
\def\gk{$(\rm G-K_{s})_{0}$}
\def\vk{$(\rm V-K_{s})_{0}$}

\begin{document}

\title{Exploiting the {\sl Gaia} EDR3 photometry to derive stellar temperatures}

\author{A. Mucciarelli\inst{1,2}, 
M. Bellazzini\inst{2}, D. Massari\inst{2,3}}

\offprints{A. Mucciarelli}
\institute{
Dipartimento di Fisica e Astronomia {\sl Augusto Righi}, Universit\`a degli Studi di Bologna, Via Gobetti 93/2, I-40129 Bologna, Italy;
\and
INAF - Osservatorio di Astrofisica e Scienza dello Spazio di Bologna, Via Gobetti 93/3, I-40129 Bologna, Italy
\and
University of Groningen, Kapteyn Astronomical Institute, NL-9747 AD Groningen, The Netherlands\\
}

\authorrunning{A. Mucciarelli et al.}
\titlerunning{\teff\  with  {\sl Gaia} EDR3}

\date{Submitted to A\&A }
 
\abstract
{We present new colour -- effective temperature (\teff ) transformations 
based on the photometry of the early  third data release  (EDR3) of the {\sl Gaia}/ESA mission. 
These relations are calibrated on a sample of about 600 dwarf and giant stars 
for which \teff\ have been previously determined with the InfraRed Flux Method from dereddened colours. 
The 1$\sigma$ dispersion of the transformations 
is of 60-80 K for the pure {\sl Gaia} colours 
\bprp\ , \bpg\ , \grp\ ,
improving to 40-60 K for colours including the 2MASS ${\rm K_s}$-band, namely $(\rm BP-K_{s})_{0}$, \rpk\ and \gk\ . 
We validate these relations in the most challenging case 
of dense stellar fields, where the  {\sl Gaia} EDR3 photometry could be less reliable, providing 
guidance for a safe use of Gaia colours in crowded environments .
We compare the \teff\ from the  {\sl Gaia} EDR3 colours with those obtained 
from standard \vk\ colours for stars in three Galactic globular clusters 
of different metallicity, namely NGC~104, NGC~6752 and NGC~7099.
The agreement between the two estimates of \teff\ is excellent, with 
mean differences between --50 and +50 K, depending on the colour, and with
1$\sigma$ dispersions around the mean \teff\ differences of 25-50 K for most of the colours and below 10 K for \bpk\ and \gk\ . This demonstrates that these colours are analogue to \vk\ , as \teff\ indicators. }

   \maketitle
%
%________________________________________________________________

\section{Introduction}

Effective temperatures (\teff ) for FGK-spectral type stars 
can be estimated with different methods either based directly on the stellar
spectra, e.g. the wings of the Balmer lines, the line-depth ratio and the excitation equilibrium, 
or on the photometric properties. 
The InfraRed Flux Method \citep[IRFM,][]{blackwell77, blackwell79, blackwell80} is one of the most popular 
methods based on photometric colours, needing accurate/precise photometry (especially 
for the infrared spectral range) and the knowledge of the colour excess, E(B-V).
Several implementations of this method have been presented in 
the literature \citep[see e.g.][]{alonso99,ramirez05,ghb09,casagrande10}. 
\teff\ derived with this method for suitable calibrators are also used 
to obtain relations between different broad-band colours and 
${\rm T_{\rm eff}}$, thus enabling an immediate estimate of \teff\ even for stars for which
the IRFM cannot be directly used.

The ESA/{\sl Gaia} mission \citep{Gaia16} is providing accurate and precise all-sky photometry 
in three broad-band photometric filters, named G, BP and RP.  
 {\sl Gaia} DR2 colour -- \teff\ transformations calibrated on IRFM ${\rm T_{\rm eff}}$ have been 
 presented by \citet[][MB20 hereafter]{mbel20} and \citet{casagrande21}.

The recent Gaia Early Data Release 3 \citep[EDR3,][]{brown20} 
has significantly improved upon the previous DR2,
including astrometric and photometric information for about 1.5 billions of stars. 
The superior quality of  {\sl Gaia} EDR3 photometry and its internal homogeneity  \citep[][R20, hereafter]{yang21,riello20}
guarantees a further improvement in the determination of stellar parameters. 
In this paper we present new colour-\teff\ 
transformations based on  {\sl Gaia} EDR3 and 2MASS photometry, validating these relations 
in the case of crowded stellar fields.

\section{New colours-\teff\ transformations}
\label{transform}

Following the same procedure adopted in MB20, we derived colours-\teff\ transformations 
for different broad-band colours including the {\sl Gaia} passbands. We used the IRFM \teff\ 
computed by \citet{ghb09} for a sample of about 450 dwarf stars (log~g$>$3.0) and 
about 200 giant stars (log~g<3.0) with metallicity between [Fe/H]$\sim$--4.0 and 0.0 dex. 
The broad-band colours that we considered in the analysis are  \bprp\ , \bpg\ , \grp\ , \gk\ , \bpk\ and \rpk\ . 
They have been derived adopting the  {\sl Gaia} EDR3 photometry \citep{brown20} 
 and the ${\rm K_{s}}$-band magnitudes from the Two Micron All Sky Survey \citep[2MASS,][]{skrutskie06}. 
 {\sl Gaia} magnitudes have been corrected for interstellar reddening following the iterative procedure 
 described in \citet{babusiaux18}, while ${\rm K_{s}}$ magnitudes have been corrected adopting 
 the extinction coefficient by \citet{mccall04}.
 Colour excess values E(B-V) are the same used by \citet{ghb09}.

We computed the best polynomial fit relating each colour C with $\theta$ (defined as $\theta$=5040/\teff\ )
and the stellar metallicity [Fe/H], according to the functional form:

\begin{equation}
\theta = {\rm b_0}+{\rm b_1}{\rm C}+{\rm b_2}{\rm C^2}+{\rm b_3}{\rm [Fe/H]}+{\rm b_4}{\rm [Fe/H]^2}+{\rm b_5}{\rm [Fe/H]}{\rm C}
\label{EQ1}
\end{equation}

and considering separately dwarf and giant stars. A few outliers have been removed adopting an iterative 
2.5$\sigma$-clipping procedure.
Table 1 lists the colour range of validity, the number of stars used 
for the fit, the 1$\sigma$ dispersion of the fit residuals and the coefficients $b_0$,...,$b_5$, 
for both dwarf and giant stars samples.

The colour-\teff\ relations that we obtained in this way have typical 1$\sigma$ dispersion of $\sim$40-60 K and $\sim$40-80 K, for dwarf and giant stars respectively. 
The 1$\sigma$ dispersion of the relations is usually adopted as a conservative estimate of the uncertainty in the 
derived \teff\, when this kind of colour-\teff\ relations are provided and/or used \citep[see e.g.][]{alonso99,ghb09,casagrande21}. 
This uncertainty should be added in quadrature to that obtained by propagating the colour error. The uncertainty in [Fe/H] has a negligible impact on the derived \teff\ , 
as a variation of $\pm$0.1 dex leads to a change in \teff\ smaller than $\sim$10 K, depending on the adopted relation. Finally, we checked that the temperature differences 
given by the relations for dwarfs and giants at the adopted dwarf/giant threshold (log~g$=$3.0) is about 10-20 K, significantly smaller that the uncertainties. 

In the common practice of abundance analysis a full propagation of the errors, including errors in the relation coefficients, is not adopted 
(we are not aware of a single example in the literature, in the field of stellar populations studies). 
The uncertainties involved in the whole process of abundance estimate are 
so many and so deeply entangled that a full propagation can be prone to underestimates 
of the actual errors on the abundances.
However, for application cases requiring full error propagation on the final \teff\ estimates, in Appendix B we provide (a) alternative relations adopting differences w.r.t. the mean colour as independent variable (e.g, using \bprp\-
$<$\bprp$>$, instead of \bprp\ alone) to minimise the off-diagonal terms of the covariance matrix, and (b) the full covariance matrices for all the relations.

The new transformations are very similar to those provided by MB20, based on {\sl Gaia} DR2 photometry, 
reflecting the similarity between the DR2 and EDR3 photometric systems. 
The use of the old relations with  {\sl Gaia} EDR3 photometry provides \teff\ that differ less than 40-50 K 
from those obtained with the new relations.
Also, the new transformations have 1$\sigma$ dispersion similar or smaller than those 
obtained with DR2 photometry.
In particular, we noted that dispersion of all of the transformations including the G-band magnitudes 
are reduced by $\sim$20-30\% with respect to those obtained with {\sl Gaia} DR2 photometry. 
Indeed, according to R20, most significant improvements between DR2 and EDR3 photometry occurred in the bright star regime
that is spanned by our calibrating sources  (G$<$6.0).

Figures A.1-A.6 in the Appendix A show the colour -- \teff\ trends for
the adopted calibrating sample and the corresponding polynomial fit.
The stars are coloured according to the metallicity interval they belong to: [Fe/H]$\leq$--2.5 dex 
(blue points),--2.5$<$[Fe/H]$\leq$--1.5 (green points), --1.5$<$[Fe/H]$\leq$--0.5 (red points), 
[Fe/H]$>$--0.5 dex (black points).
Finally, Figure A.7 shows the behaviour of the fit residuals as a function of [Fe/H] for 
all the relations.

\begin{table*}[htbp]
\caption{Coefficients ${\rm b_{0}}$,...,${\rm b_{5}}$ of the colour-\teff relations based on  {\sl Gaia} EDR3 magnitudes, 
together with corresponding colour range, the dispersion of the fit residuals and the number of used stars.}             % title of Table
\label{tab1}      % is used to refer this table in the text
\centering                          % used for centering table
\begin{tabular}{c c c c   c c c  c c c}        % centered columns (4 columns)
\hline\hline                 % inserts double horizontal lines
{\rm Colour} & {\rm Colour range} & $\sigma_{\rm T_{\rm eff}}$  & {\rm N} & 
    ${\rm b_0}$ & ${\rm b_1}$ & ${\rm b_2}$ & ${\rm b_3}$ & 
    ${\rm b_4}$ & ${\rm b_5}$  \\ 
\hline
	  &  (mag)  &   (K)   &   &    &   &      &     &      &    \\
\hline
{\bf Dwarf stars}	  &    &      &   &  &   &      &     &      &       \\
\hline

\bprp\  &    [0.39--1.50]  &  61  &   436  &  0.4929   &    0.5092    &  -0.0353    &	0.0192    &  -0.0020	&  -0.0395  \\
\bpg\   &    [0.13--0.69]  &  58  &   418  &  0.5316   &    1.2452    &  -0.4677    &	0.0068    &  -0.0031	&  -0.0752  \\
\grp\   &    [0.25--0.81]  &  62  &   439  &  0.5050   &    0.6532    &   0.2284    &	0.0260    &  -0.0011	&  -0.0726  \\
\bpk\   &    [0.62--3.21]  &  44  &   439  &  0.5342   &    0.2044    &  -0.0021    &	0.0276    &   0.0005	&  -0.0158  \\
\rpk\   &    [0.34--1.74]  &  53  &   435  &  0.5526   &    0.3712    &  -0.0121    &	0.0330    &   0.0029	&  -0.0220  \\
\gk\    &    [0.53--2.54]  &  48  &   443  &  0.5351   &    0.2440    &   0.0016    &	0.0289    &   0.0015	&  -0.0163  \\

\hline
{\bf Giant stars}	  &    &      &   &   &   &      &     &      &       \\
\hline

\bprp\  &    [0.33--1.81]  & 83  &  209  &    0.5323  &  0.4775  & -0.0344 &  -0.0110 &  -0.0020  & -0.0009 \\
\bpg\   &    [0.11--0.89]  & 83  &  208  &    0.5701  &  1.1188  & -0.3710 &  -0.0236 &  -0.0039  &  0.0070 \\
\grp\   &    [0.22--0.92]  & 71  &  201  &    0.5472  &  0.5914  &  0.2347 &  -0.0119 &  -0.0012  &  0.0060 \\
\bpk\   &    [0.68--3.97]  & 49  &  211  &    0.5668  &  0.1890  & -0.0017 &   0.0065 &  -0.0008  & -0.0045 \\
\rpk\   &    [0.35--2.23]  & 61  &  215  &    0.5774  &  0.3637  & -0.0226 &   0.0346 &   0.0007  & -0.0221 \\
\gk\    &    [0.57--3.10]  & 46  &  206  &    0.5569  &  0.2436  & -0.0035 &   0.0211 &   0.0007  & -0.0089 \\

\hline                  
\hline                                   %inserts single line
\end{tabular}
\end{table*}

We compared the predictions of the \citet{casagrande21} relations with ours for the stars of our calibrating sample, 
using all the colours that can be obtained  combining the three {\sl Gaia} pass-bands among them and with 2MASS K. 
The mean differences are within $\simeq \pm 100$~K and the scatter is small (spanning $\la 50$~K) 
in all cases except for the \bpg\ colour, where dwarfs display a mean difference of about 250~K and 
a significant scatter ($\ga 100$~K). Taking into account that part of the observed differences may be due to 
the subtle changes between {\sl Gaia} DR2 and EDR3 photometry (especially for G$\le 13.0$, see \citet{dafydd}, R20), 
we can conclude that the two calibrations provide consistent results, within the uncertainties. 
The \citet{casagrande21} relations use fourteen coefficients and explicitly include the dependency 
on surface gravity, hence they may be appropriate when all the astrophysical parameters of the target stars, 
except \teff\, are already known with high accuracy. On the other hand, our relations accounts for 
the tiny effect of surface gravity by means of a simple giants/dwarfs dichotomy and are defined by 
just five parameters. They are simpler and have a wider range of applicability in most real cases.

\section{Application on three globular clusters: NGC~104, NGC~6752, M30}

The new relations are based on isolated bright field stars for which {\sl Gaia} provides superb photometry, 
usually not affected by issues related to stellar contamination 
and/or background subtraction.
To test the effectiveness of our relations in determining reliable \teff\ in any condition, 
we decided to validate them in dense stellar fields, where the superior photometric quality 
of the {\sl Gaia} magnitudes can be hampered by the high stellar crowding.

The selected stellar fields to perform such a test correspond to three Galactic globular clusters (GC), 
namely NGC~104 (47 Tucanae), NGC~6752 and NGC~7099 (M30). 
These were selected according to the following criteria:

\begin{enumerate}
\item{} span the entire range of metallicity covered by the population of Galactic clusters, 
with the selection of a metal-rich GC [NGC~104, Fe/H]=--0.75 dex) 
a metal-intermediate GC (NGC~6752, [Fe/H]=--1.49 dex) 
and a metal-poor GC (NGC~7099, [Fe/H]=--2.31 dex), according to the iron abundances derived by \citet{mbon20}. 
The reason behind the choice of clusters with different [Fe/H] is to check 
the validity of our transformations against the metallicity, because 
this parameter enters our Eq.~\ref{EQ1} directly.
\item{} low colour excess E(B-V) \citep[between 0.04 and 0.07 mag, see][]{mbon20}, to 
minimise the effect of uncertainties in the extinction on the derivation of \teff\ .
\item{} GCs having available  ground-based V photometry from 
the database maintained by P. B. Stetson \citep{stetson19} 
and and ${\rm K_{s}}$-band photometry from 2MASS \citet{skrutskie06}.
This is to derive a reference \teff\ by using homogeneous \vk\ colours.
\end{enumerate}

 Clusters members were first selected to have proper motions within 1.5 mas/yr (for NGC~104 and NGC~6752) 
 and 1.0 mas/yr (for NGC~7099) from the cluster mean proper motions as given by \citet{baumgardt19}.  
Then we filtered stars based on "goodness of measure" EDR3 quality parameters, 
following prescriptions provided by \citet{lindegren18} and R20, including in our final samples
only stars having: \\
{\sl (i)} ruwe$<$1.4;\\
{\sl (ii)}  $\mid C^{*} \mid < 2 \sigma_{c}$, where 
$C^{*}$ and $\sigma_{c}$ are defined according to Eq.~6 and 18, respectively, in R20.

For these cluster stars we computed \teff\ 
adopting the six colour-\teff\ transformations derived in Section~\ref{transform}.
Additionally, reference \teff\ have been computed using the \vk\ -- \teff\ transformation by \citet{ghb09}. 
The latter is based on the same sample of stars and IRFM \teff\ used to derived our own relations, hence all these 
\teff\ are  on the same scale.
We restricted this analysis to the stars with G$<$17 
and with error in \vk\ smaller than 0.03 mag
in order to exclude stars with large uncertainties in the 2MASS ${\rm K_{s}}$ magnitudes.
To be sure that the different fitting procedures used here for the {\sl Gaia} colours and 
used by \citet{ghb09} for \vk\  do not introduce systematics in the computed \teff\ , we 
derived the \vk\ -- \teff\ transformation adopting our procedure and the \vk\ already used 
by \citet{ghb09}. The average difference of the \teff\ from the two \vk\ -- \teff\ transformations 
for the cluster stars is of +1 K ($\sigma$=~6 K). Hence, our fitting procedure does not introduce 
differences with respect to the transformations by \citet{ghb09} and we can compare \teff\ 
from {\sl Gaia} and \vk\ colours.

\begin{figure*}
\centering
\includegraphics[width=0.9\textwidth]{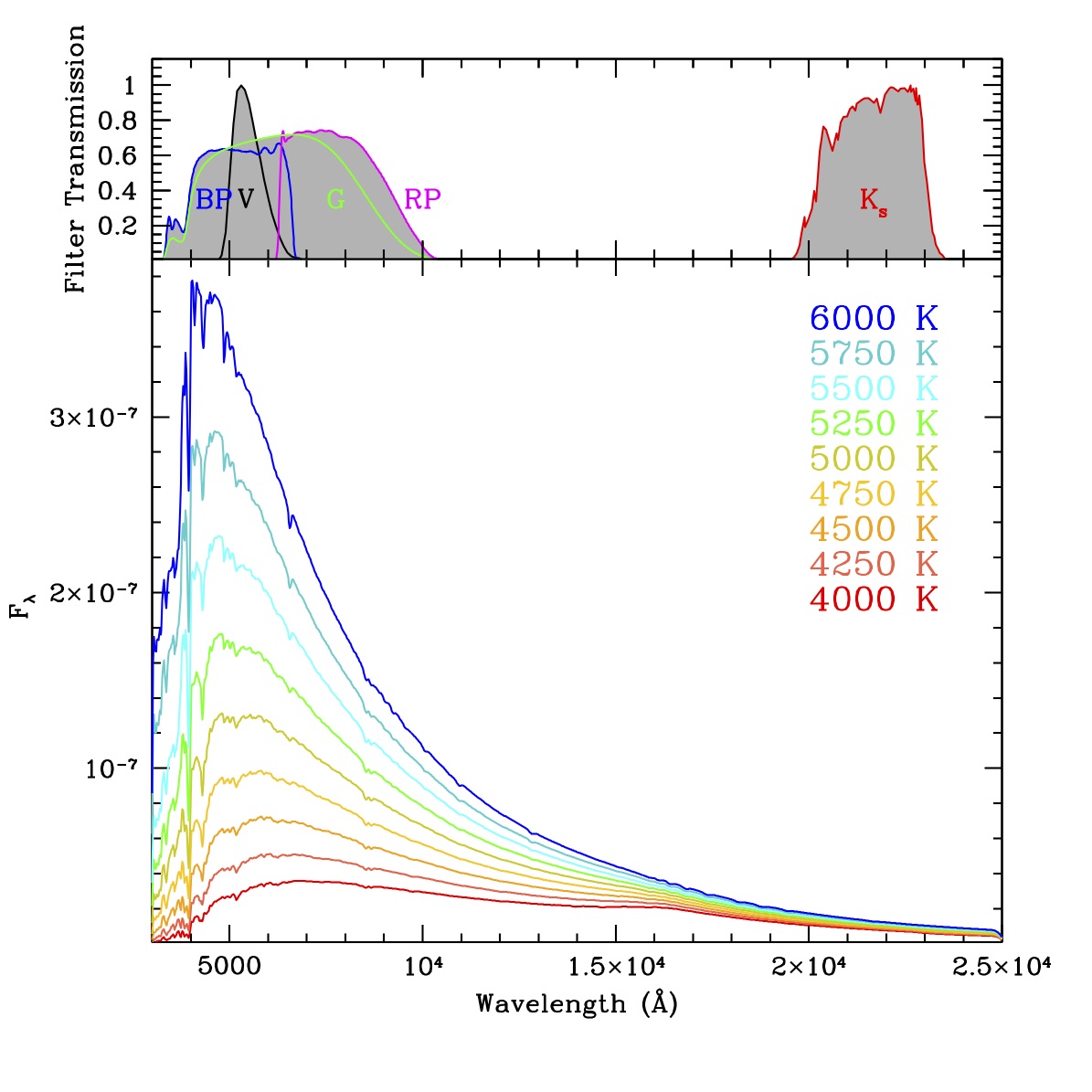}
\caption{Main panel: synthetic spectra calculated with \teff\ from 4000 K (spectrum with the lower flux) 
to 6000 K (spectrum with the higher flux) with step of 250 K. All the spectra adopt [M/H]=--1.0 dex. The upper panel 
shows the profile of the photometric filters used in this 
work.}
\label{filter}
\end{figure*}

From the results of our analysis, it is clear that each photometric colour has pros and cons 
as \teff\ indicator, due to different 
wavelength baseline, sensitivity to \teff\  and to other parameters, i.e. metallicity and surface gravity.
Here we adopt \teff\ from the \vk\ colour as reference values to check the robustness of those derived 
from  {\sl Gaia} EDR3 photometry. The \vk\ colour is one of the most common and reliable photometric indicator of \teff\ 
\citep[see e.g.][]{fernley89,bessel98,alonso99}
because of two main factors:\\
{\sl (i)}~the different sensitivity to \teff\ of the flux in V- and ${\rm K_s}$-bands. 
This effect is clearly visible in Fig.~\ref{filter} that shows a set of synthetic fluxes
with \teff\ from 4000 to 6000 K with step of 250 K.
All the synthetic spectra have been calculated with 
the {\tt SYNTHE} spectral synthesis code \citep{kurucz05}.
The V-band flux is highly sensitive to \teff\ , increasing by a factor of 10 as \teff\ ranges 
from 4000 to 6000 K. On the other hand, the ${\rm K_s}$-band flux increases by only a factor of 1.5 
in spite of the same \teff\ change. Therefore, \vk\ effectively behaves as the ratio between a \teff\--sensitive flux 
and an almost \teff\--insensitive flux.\\
{\sl (ii)}~the Johnson-Cousins V band photometry is better standardised
than other optical bands, like B and I bands, whose definitions 
can vary depending on the adopted photometric system \citep{bessel88}.

%%%%%%%%%%%%%%%%%%%%%%%

\begin{figure*}
\centering
\includegraphics[width=1.4\columnwidth]{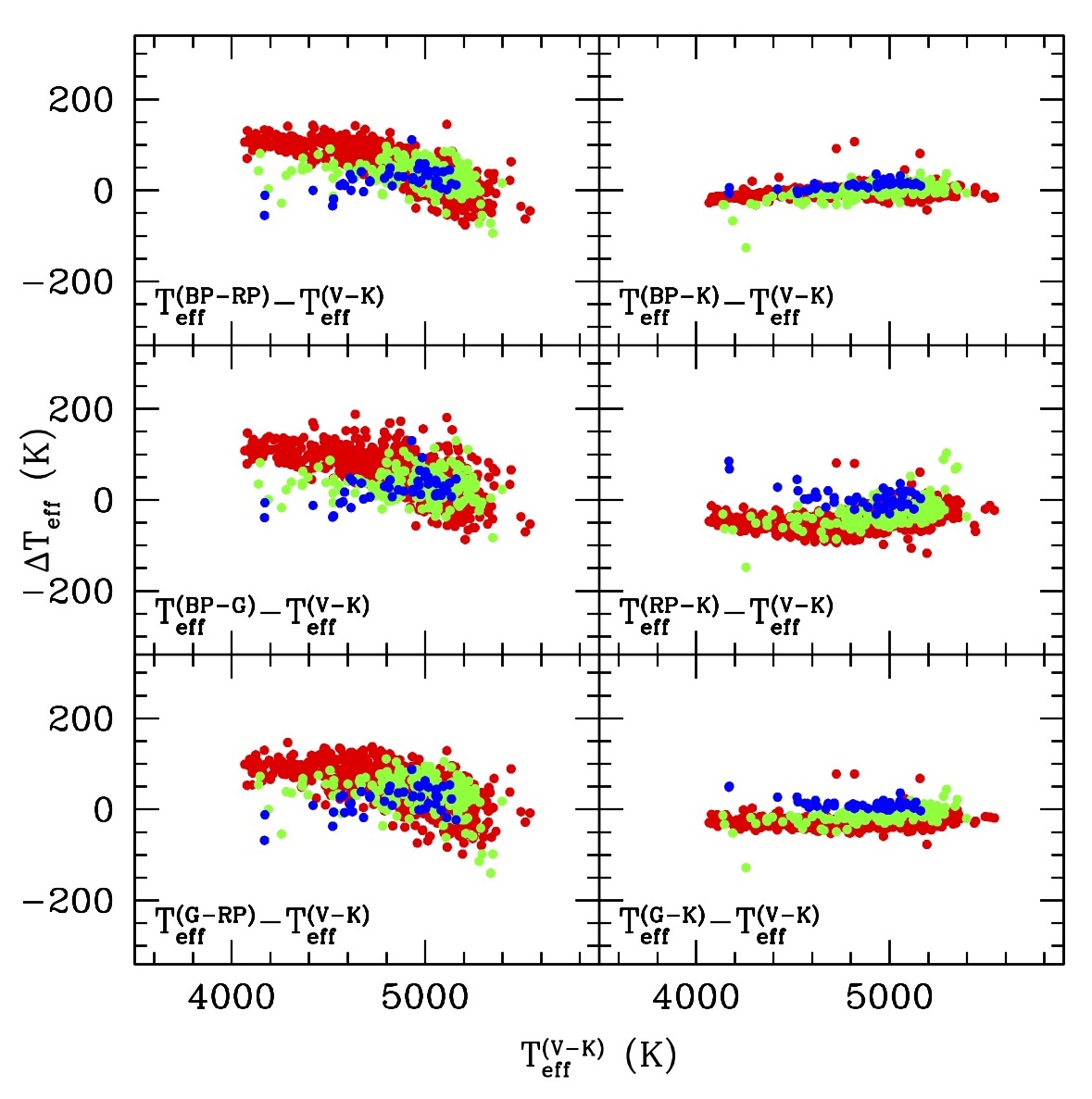}
\caption{Differences between \teff\ derived from the {\sl Gaia} colours 
and from \vk\ as a function of the \vk\-based \teff\ for the giant stars in three Galactic globular clusters, namely NGC~7099 (blue points), NGC~6752 (green points) and NGC~104 (red points).}
\label{deltat}
\end{figure*}

Fig.~\ref{deltat} shows the behaviour of the differences between 
\teff\ from  {\sl Gaia} EDR3 colours and from \vk\
as a function of the latter for the stars in the three GCs.
The average values of these differences,  the corresponding 1$\sigma$ dispersion and 
number of stars are listed in Table 2.

\teff\ from pure {\sl Gaia} EDR3 colours have mean differences with respect to the reference \teff\ 
between +20 and +50 K, with scatter between 25 and 50 K.  
In the case of the metal-rich GC NGC~104 a mild trend of $\Delta$\teff\ 
with the reference \teff exist, in the sense that the  {\sl Gaia} EDR3 \teff\ becomes slightly hotter than those from \vk\ for the coldest stars. 

The colours including ${\rm K_s}$ magnitudes have small average differences and 1$\sigma$ dispersions , 
in particular \bpk\ and \gk\ provide the best agreement with the \teff\ from \vk\ , 
with 1$\sigma$ dispersions smaller than 10 K. 
This simple test demonstrates that:
\begin{itemize}
 \item{} ({\sl Gaia}-${\rm K_s}$) colours are analogue to \vk\ as \teff\ indicators because they  
have a large wavelength baseline including filters with different sensitivity in \teff\   
\item{}  the calibrations based on field (isolated) stars work well also in crowding conditions typical of nearby Galactic GCs (D$\la 10.0$~kpc), once the simple selections based on quality parameters described above are adopted (see below, for further discussion). 
\end{itemize}

\begin{table*}[htbp]
\caption{Average differences between \teff\ derived from the {\sl Gaia} EDR3 colours and \vk\ for the globular clusters NGC~104, NGC~6752 and NGC~7099, together with the 1$\sigma$ dispersion and the number of used stars.}             % title of Table
\label{tab2}      % is used to refer this table in the text
\centering                          % used for centering table
\begin{tabular}{c c c c   c c c  c c c}        % centered columns (4 columns)
\hline\hline                 % inserts double horizontal lines
 $\Delta$\teff\	 &   & NGC~104   &   &  & NGC~6752 & &  & NGC~7099 &  \\ 
\hline
   & $<\Delta$\teff\ $>$  &   $\sigma$   &  ${\rm N_{star}}$ &   $<\Delta$\teff\ $>$   & $\sigma$     &  ${\rm N_{star}}$       & $<\Delta$\teff\ $>$   &  $\sigma$        &   ${\rm N_{star}}$    \\
\hline
	  &  (K)   &   (K)   &   &  (K)   & (K)  &      &   (K)  &  (K)    &       \\
\hline

\bprp\ - \vk\ &   +48   &  45  &  699   &  +35   &   26     &  185  &	+24   & 18	&    41 \\
\bpg\ - \vk\  &   +54   &  48  &  697   &  +31   &   28     &  185  &	+19   & 25	&    42 \\
\grp\ - \vk\  &   +51   &  41  &  685   &  +41   &   24     &  178  & 	+20   & 24	&    42 \\
\bpk\ - \vk\  &   --3   &   7  &  671   &   +2   &    8     &  171  &	+11   &  8	&    43 \\
\rpk\ - \vk\  &  --45   &  19  &  686   & --31   &   15     &  178  &	 +0   & 15	&    41 \\
\gk\ - \vk\   &  --25   &   8  &  684   & --10   &    9     &  179  &	 +9   &  7	&    41 \\

\hline                  
\hline                                   %inserts single line
\end{tabular}
\end{table*}

\section{How to derive accurate \teff\ in crowded stellar fields}
The use of the  {\sl Gaia} EDR3 photometry to infer \teff\ in dense stellar fields 
(like globular clusters) needs a note of caution because the {\sl Gaia} magnitudes, 
regardless of their formal small uncertainties, can be affected by issues 
concerning the background subtraction and contamination by neighbouring stars (R20).

\begin{figure*}[ht]
\centering
\includegraphics[width=0.8\textwidth]{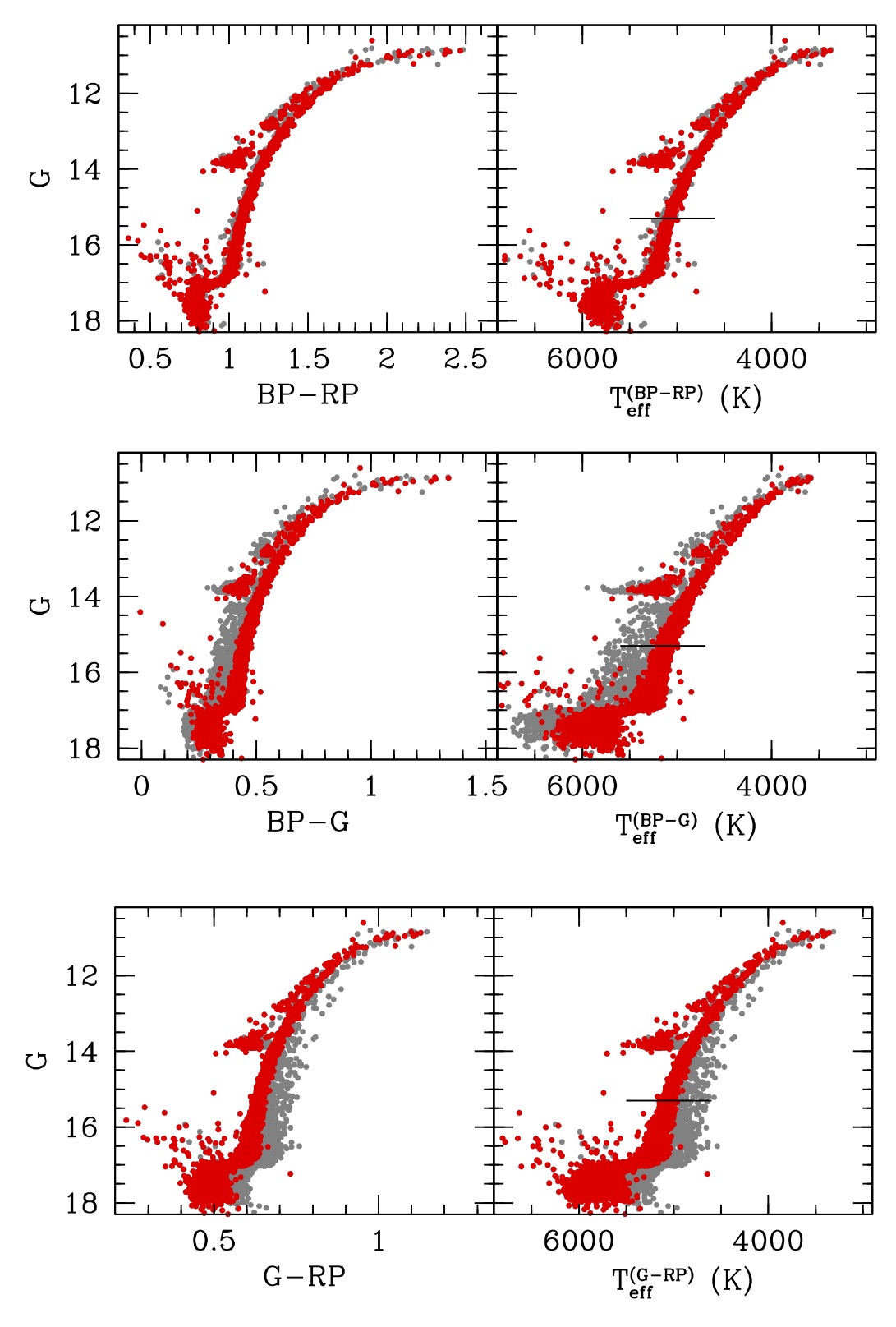}
\caption{Colour-magnitude diagrams for NGC~104 considering the pure {\sl Gaia} colours (left panels) and the corresponding \teff\ vs G-band magnitudes diagrams (right panels). Red and grey points mark
the stars selected and rejected according to the criterion provided by R20, respectively. 
The horizontal lines in the right panels mark the transition between the dwarf and giant stars regimes.}
\label{47tuc_sel}
\end{figure*}

Left panels of Fig.~\ref{47tuc_sel} show the three colour-magnitude 
diagrams for NGC~104 including pure {\sl Gaia} colours. 
At variance with \bprp\ , the other two colours 
show an asymmetric broadening of the red giant branch (RGB) that becomes 
more evident for G$>$14. 
In particular, an excess of stars bluer than the 
main locus of the RGB is visible when we use \bpg\ , 
while with \grp\ the situation is the opposite, with an 
excess of redder stars. 
Obviously, these anomalous colours translate in 
anomalous \teff\ that can be as discrepant as $\pm$500 K 
compared to RGB stars with the same G magnitude. 
 Indeed, BP and RP magnitudes are known to be more prone
to contamination from light not related to the target sources (e.g., nearby stars)
than G magnitudes, for reason inherent with the different way in which BP/RP and G fluxes are 
acquired and processed (R20).

Stars with anomalous colours can be easily identified and excluded 
by applying the criterion used in Section 2 for the three 
target clusters ($\mid C^{*} \mid < 2 \sigma_{c}$).
In Fig.~\ref{47tuc_sel} the sources fulfilling this 
criterion (therefore considered as high-quality/reliable photometry sources) are 
shown as red circles, while the excluded stars as grey circles. 
This exercise provides three important results:
\begin{itemize}
\item the criterion $\mid C^{*} \mid < 2 \sigma_{c}$ allows to efficiently 
identify stars with possible issues related to 
background subtraction and stellar contamination;
\item this procedure is essential whether \bpg\ or \grp\ are used. In fact, 
only reliable sources provide reliable \teff\ while the other sources 
significantly over- or under-estimate 
(for \grp\ and \bpg\ , respectively) \teff\ ;
\item the symmetrical effect observed in \grp\ and \bpg\ (and in the corresponding \teff ) 
is largely cancelled out when \bprp\ is adopted. 
In fact, \bprp\ of reliable and contaminated stars provide indistinguishable \teff\ . 
\end{itemize}

In conclusion, according to this limited set of experiments, reliable \teff\ in (non extreme)
crowded fields can be obtained by selecting out stars with corrupted colours with criteria
based on quality parameters provided in the Gaia source catalogue. The criterion proposed here
($\mid C^{*} \mid < 2 \sigma_{c}$ ) is simple and very effective, in the considered cases.
However there may be cases where only stars not fulfilling such criteria are
available for the analysis. The results presented above suggest that reliable \teff\ estimates can
obtained also for these stars, using \bprp\  as \teff\ indicator, taking advantage of the fact that 
BP and RP magnitudes are similarly affected by any light contamination entering the 
aperture window of BP and RP spectrophotometry (see R20, for discussion).

\section{The impact of the {\sl Gaia} \teff\ on the chemical abundances}

As a sanity check, we evaluated the impact of the new colour-\teff\ transformations 
derived from  {\sl Gaia} EDR3 photometry on the chemical abundances from high-resolution spectra. 
We consider the data-set of high-resolution spectra acquired with the spectrograph UVES at the Very Large 
telescope of ESO for giant stars in 16 Galactic GCs already analysed in \citet{mbon20}.
The iron abundance of these stars have been derived 
following the same procedure adopted by \citet{mbon20} and
using the new \teff\ scales. These new [Fe/H] values have 
been compared with those obtained for the same stars from \vk\ 
by \citet{mbon20}.

\begin{figure}[h]
\centering
\includegraphics[width=\columnwidth]{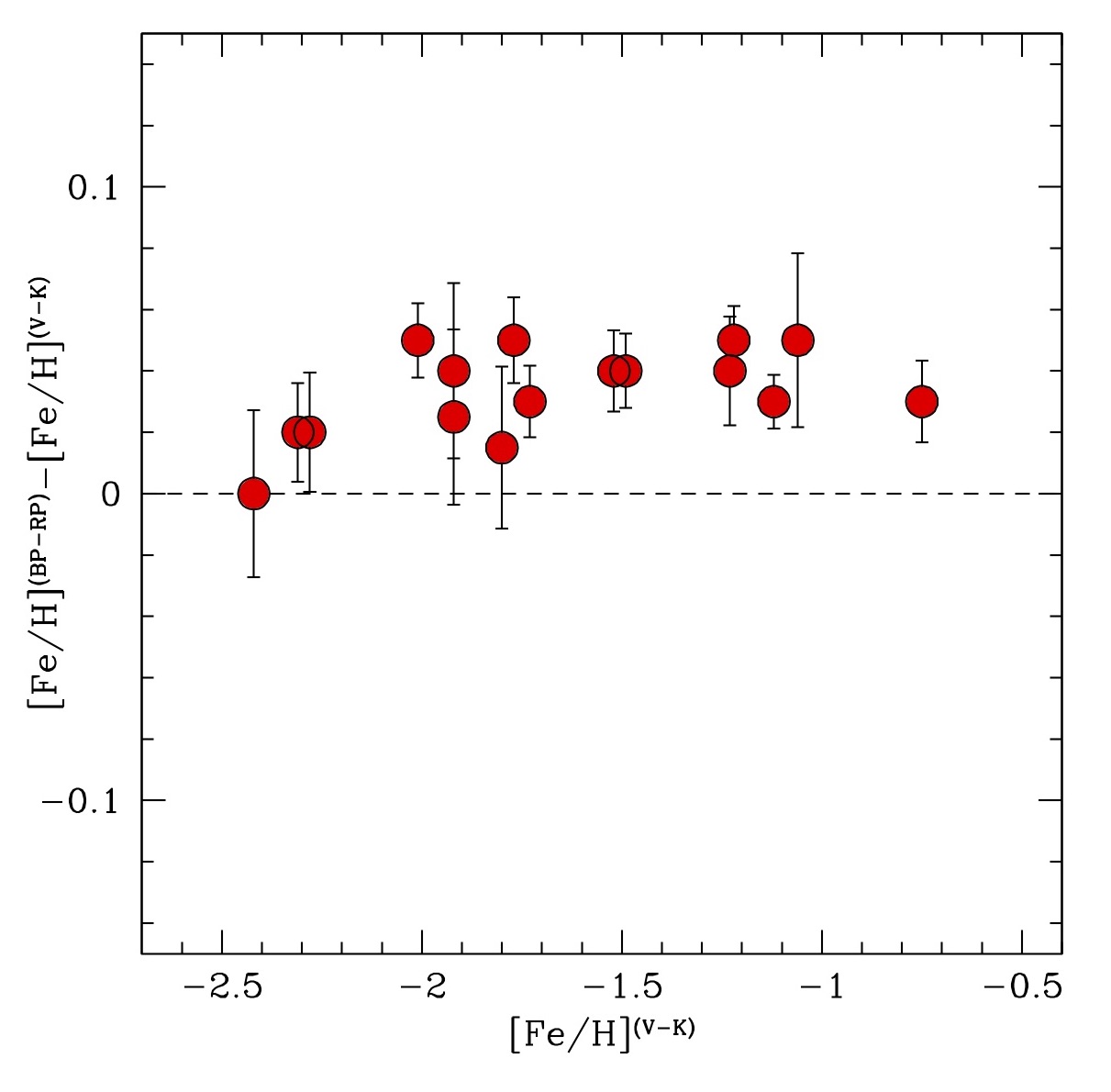}
\caption{Behaviour of \vk\- and \bprp\-based [Fe/H] as a function 
of the iron content [Fe/H] derived from \vk\-based \teff\ 
for the 16 Galactic globular clusters analysed in \citet{mbon20}. 
}
\label{delta}
\end{figure}

In the case of pure {\sl Gaia} colours, 
the new \teff\ are comparable with those 
from \vk\ , with differences smaller than 100 K.
These new \teff\ lead to higher [Fe/H], 
with differences between 0.01 and 0.05 dex 
with respect to the values obtained from \vk\ \teff\ .
Fig.~\ref{delta} shows, as example, the difference
of the derived [Fe/H] adopting \teff\ from \bprp\ and \vk\ . 
The average [Fe/H] difference is +0.03 dex ($\sigma$=~0.01 dex).

The {\sl Gaia} colours including ${\rm K_{s}}$ magnitudes provide a value of \teff\ for the 
spectroscopic targets which is basically indistinguishable from the one from \vk\ , 
such that the average impact in terms of [Fe/H] is smaller than 0.01 dex.

We conclude that the use of \teff\ from  {\sl Gaia} EDR3 photometry leads to 
chemical abundances fully consistent with those obtained adopting \teff\ 
from standard colours.

\section{Conclusions}
We exploited the  {\sl Gaia} EDR3 photometry to derive new colour-\teff\ transformations based on the IRFM \teff\ provided by \citet{ghb09} for a sample of about 600 bright dwarf and giant field stars. 
These transformations have typical uncertainties of 40-80 K and 40-60 K for 
giant and dwarf stars, respectively. We checked the validity of these transformations in the case of globular cluster stars, where the superior photometric quality of the {\sl Gaia} magnitudes can be hampered by the high stellar crowding, providing guidelines for safe estimates of \teff\  in these cases.
 In summary, the  {\sl Gaia} EDR3 photometry can be safely used to derive precise and accurate \teff\, with the  following recommendations:
\begin{enumerate}
\item When reliable ${\rm K_{s}}$-band photometry is available, mixed colours ({\sl Gaia} - ${\rm K_{s}}$) 
should be preferred,  as they display the maximum sensitivity to temperature. In particular \bpk\ and \gk\ are the best choice because their colour-\teff\ transformation have the smallest dispersion and show the best agreement with \teff\ derived from \vk\ .
\item When ${\rm K_{s}}$-band photometry is not available 
or not precise enough, pure {\sl Gaia} colours can be used to derive \teff\ even if with a slightly larger  dispersion with respect to the broad band colours including  ${\rm K_{s}}$.
\item BP and RP-band magnitudes in crowded fields can be affected by issues concerning stellar blending and background subtraction, despite their high photometric precision. For this reason, \grp\ and \bpg\ can lead to under- and over-estimated \teff\ , respectively. 
To avoid these effects stars should be selected according to $C^{*}$ and we recommend the criterion  $\mid C^{*} \mid < 2 \sigma_{c}$ .
Alternatively, \bprp\ should be preferred to other combinations of Gaia magnitudes, since the effects of contamination from light not related to the target source are similar in the BP and RP bands and almost cancel out when they are subtracted.
\end{enumerate}

\begin{acknowledgements}
We thank the referee, Floor van Leeuwen, for the comments and suggestions that improved the manuscript. 
We also thank Paolo Montegriffo for the useful discussions.
This work has made use of data from the European Space Agency (ESA) mission
{\it Gaia} (\url{https://www.cosmos.esa.int/gaia}), processed by the {\it Gaia}
Data Processing and Analysis Consortium (DPAC,
\url{https://www.cosmos.esa.int/web/gaia/dpac/consortium}). Funding for the DPAC
has been provided by national institutions, in particular the institutions
participating in the {\it Gaia} Multilateral Agreement.
This research has made extensive use of
the  R  programming  language  and  R  Studio environment.
\end{acknowledgements}

\bibliographystyle{apj}
{}

\clearpage
\newpage

\begin{appendix}
\section{Colour-\teff\ polynomial fits}

\begin{figure}[h]
\centering
\includegraphics[width=\columnwidth]{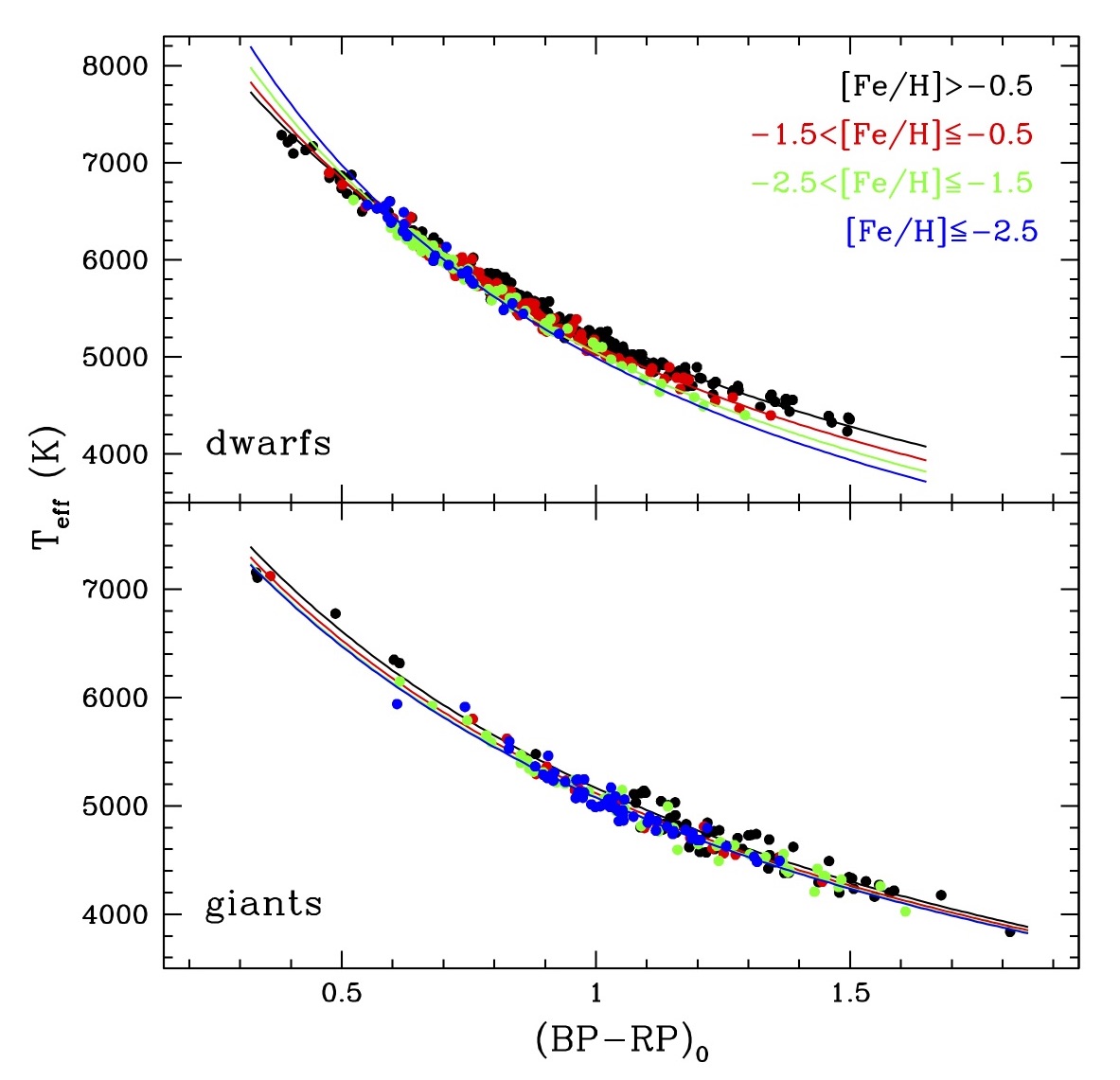}
\caption{Behaviour of \teff\ derived
from IRFM by \citet{ghb09} as a function of the 
\bprp\ colour, for dwarf and giant stars (upper and lower panels, respectively). The stars are grouped according to their metallicity: [Fe/H]$\leq$--2.5 dex 
(blue points),--2.5$<$[Fe/H]$\leq$--1.5 (green points), --1.5$<$[Fe/H]$\leq$--0.5 (red points), 
[Fe/H]$>$--0.5 dex (black points). The solid lines are the theoretical colour-\teff relation calculated with 
[Fe/H]=--3.0 dex (blue line), --2.0 dex (red line), --1.0 dex (red line), +0.0 dex (black line).}
\label{fit_bprp}
\end{figure}

%%%%%%%%%%%%%%%%%%%%%%%
\begin{figure}[h]
\includegraphics[width=\columnwidth]{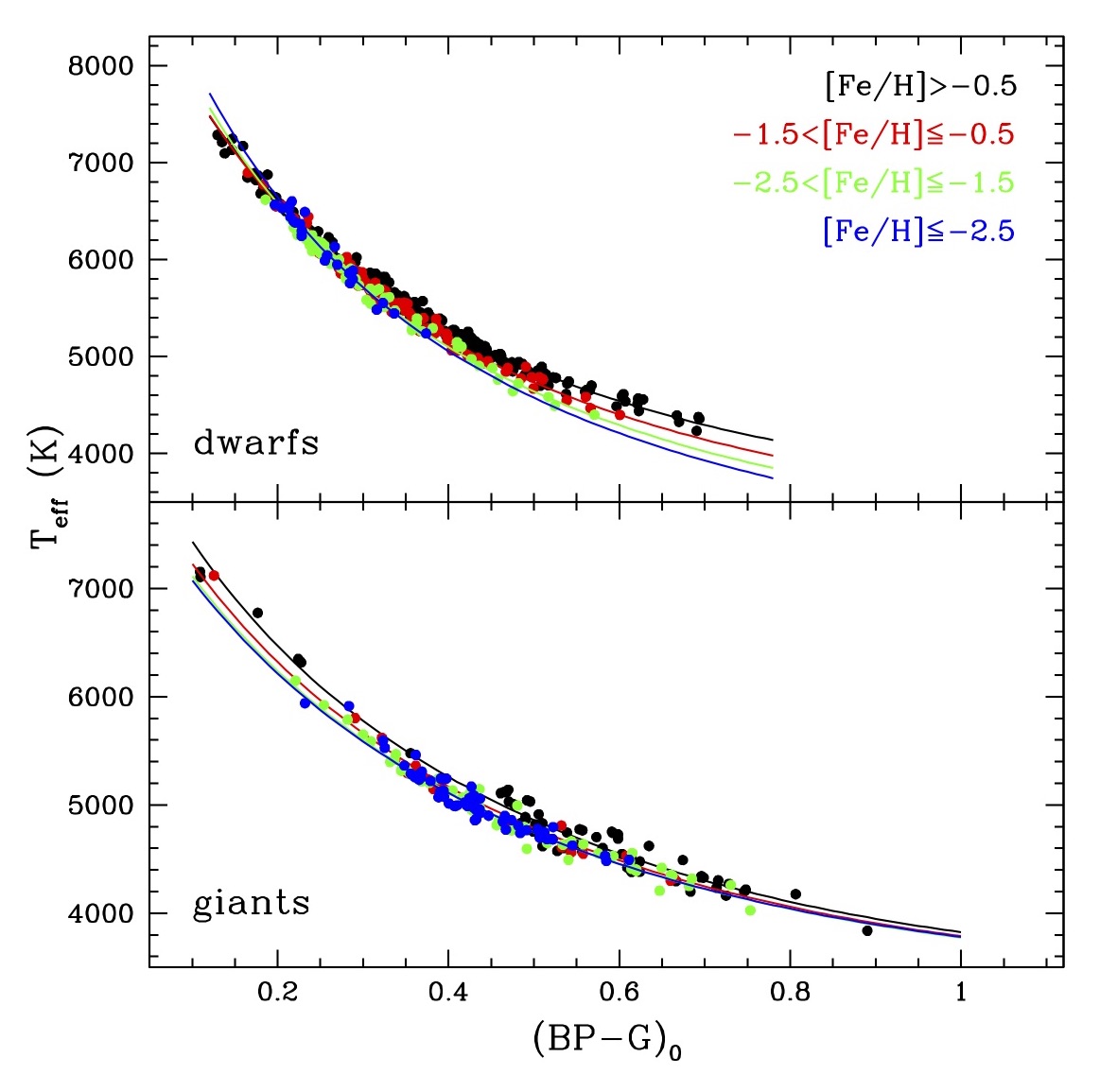}
\caption{Same of Fig.~\ref{fit_bprp} but for the  
\bpg\ colour.}
\label{bpg}
\end{figure}

%%%%%%%%%%%%%%%%%%%%%%%
\begin{figure}[h]
\includegraphics[width=\columnwidth]{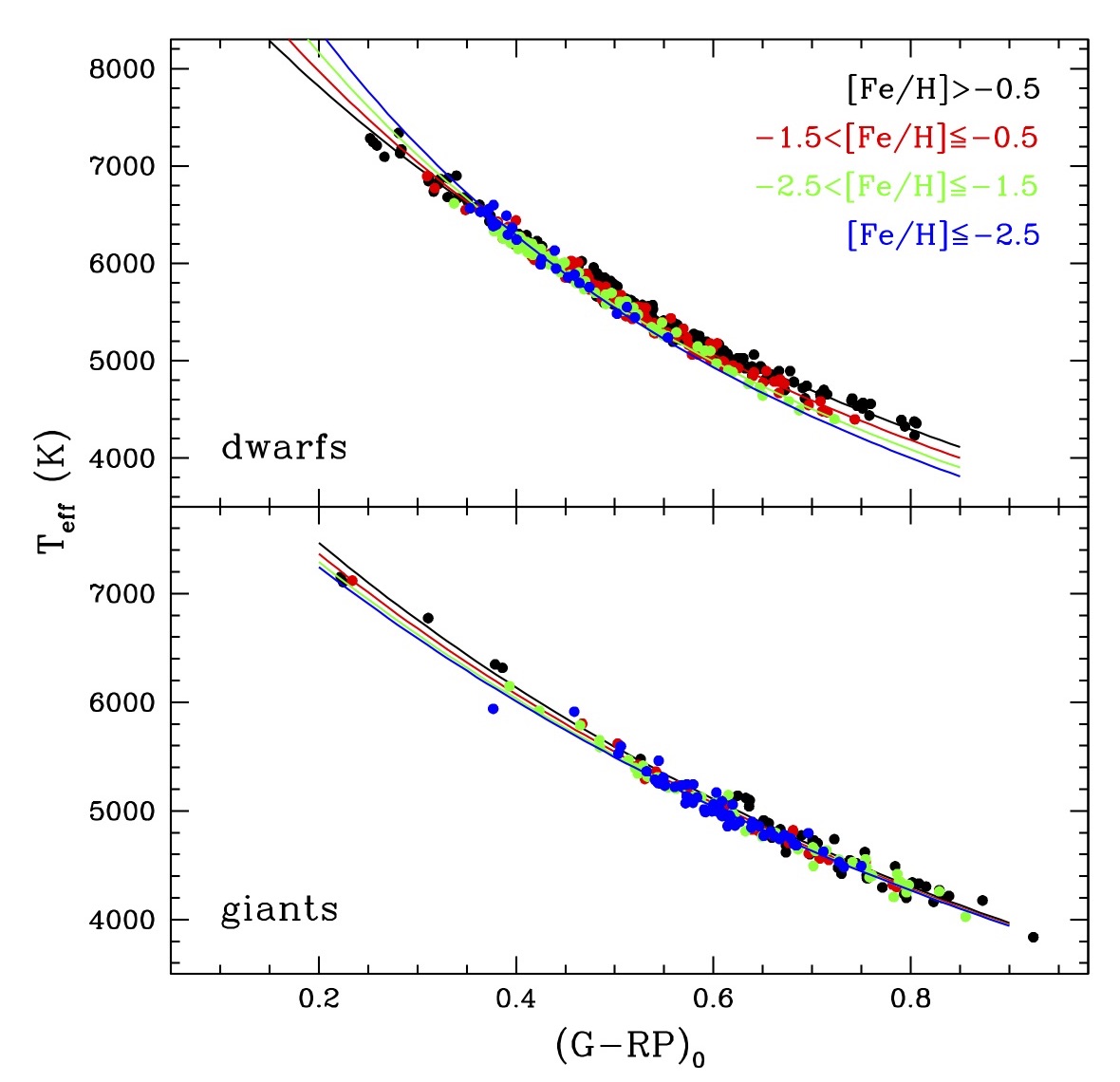}
\caption{Same of Fig.~\ref{fit_bprp} but for the  
\grp\ colour.}
\label{grp}
\end{figure}

%%%%%%%%%%%%%%%%%%%%%%%
\begin{figure}[h]
\includegraphics[width=\columnwidth]{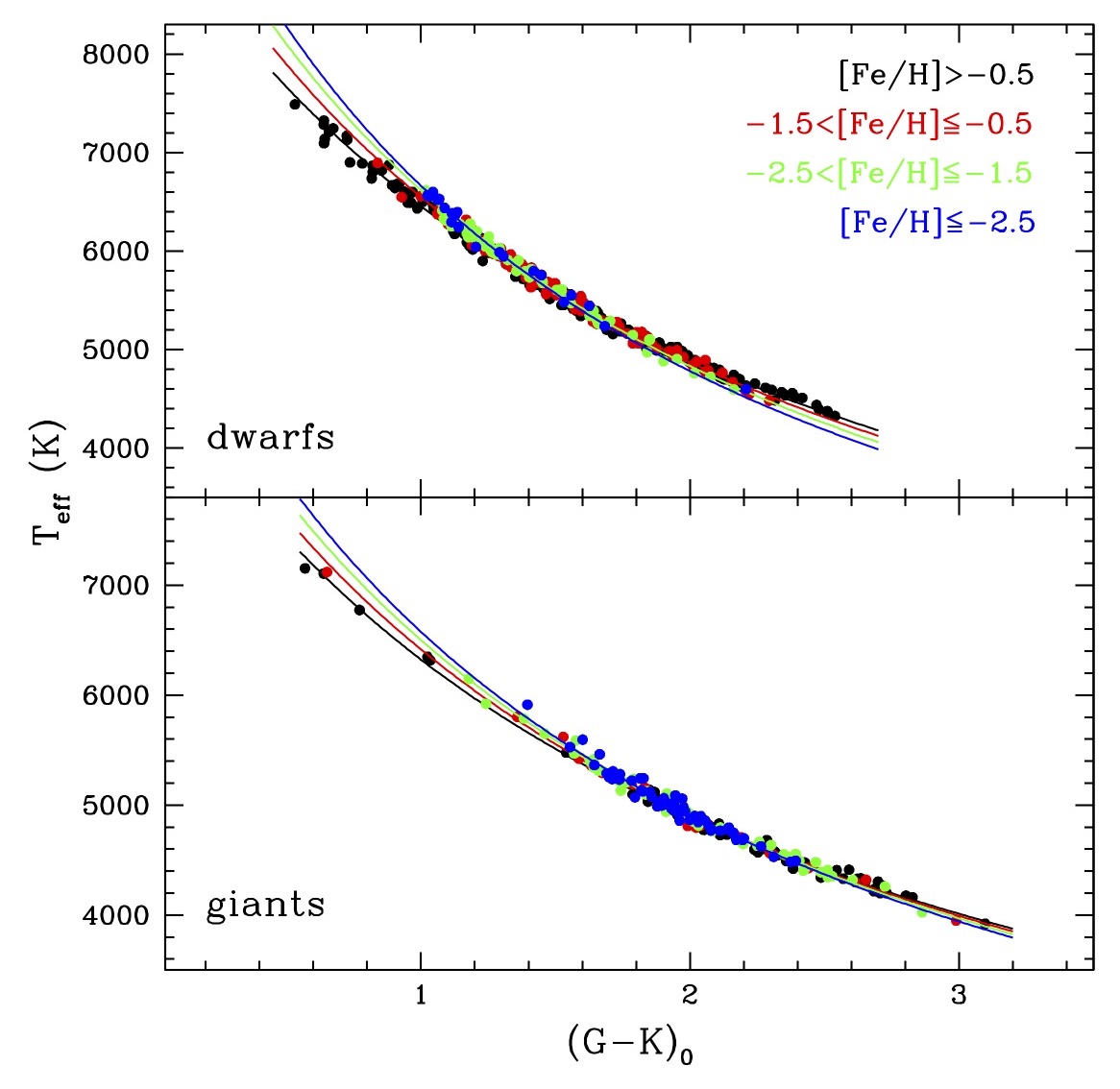}
\caption{Same of Fig.~\ref{fit_bprp} but for the  
\gk\ colour.}
\label{gk}
\end{figure}

%%%%%%%%%%%%%%%%%%%%%%%
\begin{figure}[h]
\includegraphics[width=\columnwidth]{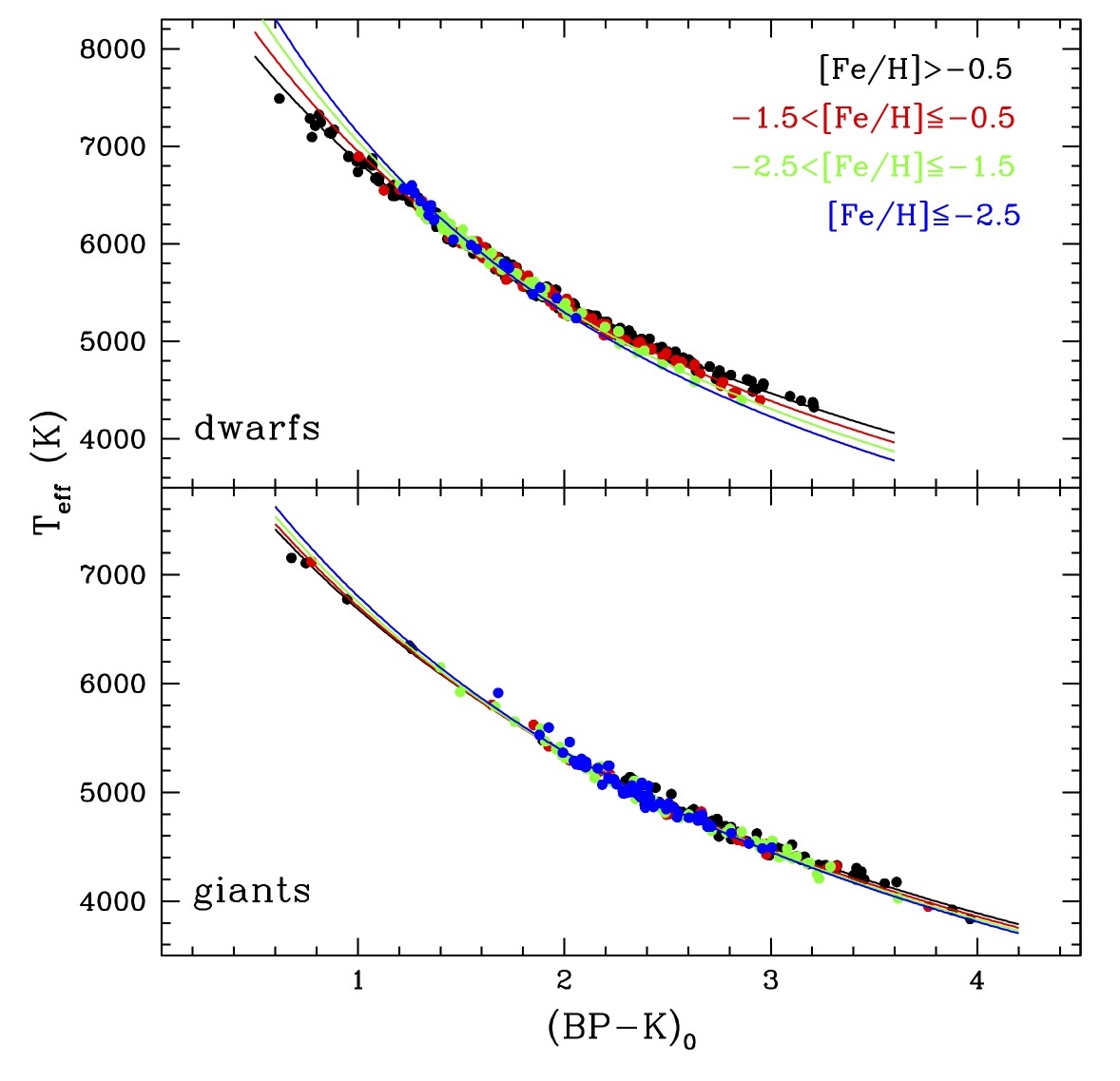}
\caption{Same of Fig.~\ref{fit_bprp} but for the  
\bpk\ colour.}
\label{bpk}
\end{figure}

%%%%%%%%%%%%%%%%%%%%%%%
\begin{figure}[h]
\includegraphics[width=\columnwidth]{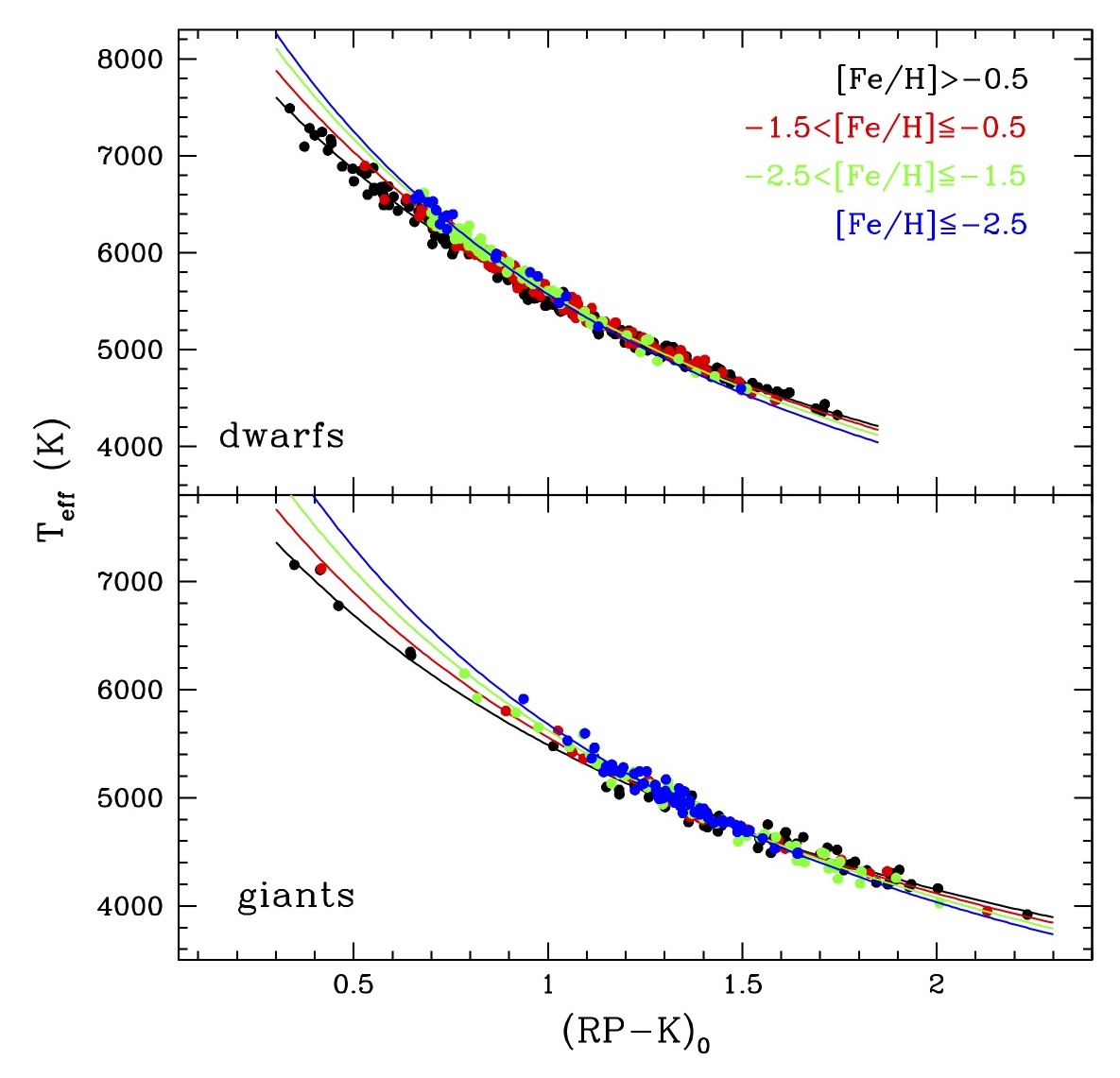}
\caption{Same of Fig.~\ref{fit_bprp} but for the  
\rpk\ colour.}
\label{rpk}
\end{figure}

%%%%%%%%%%%%%%%%%%%%%%%
\begin{figure}[h]
\includegraphics[width=\columnwidth]{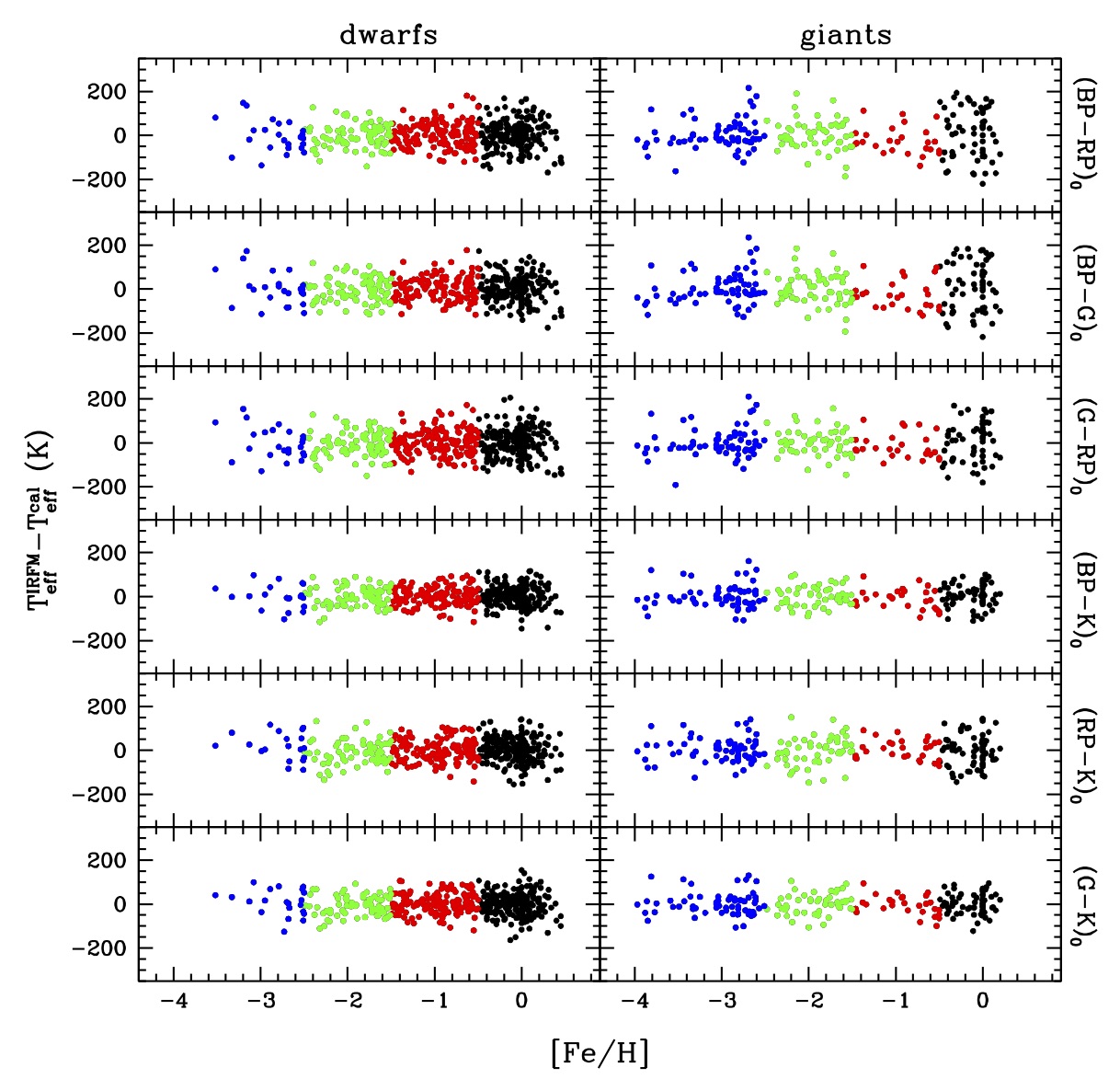}
\caption{Behaviour of the temperature residuals as a function 
of [Fe/H] for all the colour-\teff\ transformations discussed in this 
work. Colours are the same of previous figures.}
\label{resid1}
\end{figure}

\newpage
\clearpage

\section{An alternative set of colour-\teff\ transformations}
\label{alter}

As explained in Section~\ref{transform}, the usual 
approach to estimate the uncertainty in \teff\ derived 
from colour-\teff\ transformations is to propagate the 
colour error, sometimes adding in quadrature the 1$\sigma$ dispersion of the fit residuals taken 
as conservative estimate of the relation error.
An appropriate propagation of the errors, including also the uncertainties on the fit parameters and their possible covariance terms 
is nonetheless provided in the following, by means of an alternative set of colour-\teff\ transformations 
obtained with following fitting formula (reducing the off diagonal terms of the covariance matrix):

\begin{equation}
\theta = {\rm c_0}+{\rm c_1}{\rm C^*}+{\rm c_2}{\rm C^{*2}}+{\rm c_3}{\rm [Fe/H]^*}+{\rm c_4}{\rm [Fe/H]^{*2}}+{\rm c_5}{\rm [Fe/H]^*}{\rm C^*}
\label{EQ2}
\end{equation}

where ${\rm C^*}$ is the colour subtracted by the mean colour
and ${\rm [Fe/H]^*}$ is the metallicity subtracted by the mean metallicity. Table~\ref{tab2} lists the coefficients 
${\rm c_i}$ and the mean colour for each transformation 
(for all of them we assume --1.5 dex as mean metallicity). 
The 1$\sigma$ dispersion and the number of used stars are the same listed in Table~\ref{tab1}. 

For a given pair of colour/metallicity, the relations \ref{EQ1} and \ref{EQ2} (and the corresponding coefficients) {\sl provide exactly the same results}. Relation \ref{EQ1} 
is more direct to use without needing to scale both colour and metallicity to the mean values used of the calibrators sample. 
It can be used when the 1$\sigma$ dispersion of the fit is assumed as reliable estimate of the \teff\ error due to the 
calibration itself.
Relation \ref{EQ2} needs the scaling of both colour and metallicity to the mean values used of the calibrators sample and should be used when the user is interested in calculating 
the \teff\ uncertainty by propagating also the errors in the coefficients.

We list the  normalised covariance matrix for each transformation. 
In each matrix, the raws and the columns correspond to the parameters ${\rm c_0}$, ${\rm c_1}$, ${\rm c_2}$, ${\rm c_3}$, ${\rm c_4}$ and  ${\rm c_5}$, in this order.

\begin{table*}[htbp]
\caption{Coefficients ${\rm c_{0}}$,...,${\rm c_{5}}$ of the colour-\teff relations (see Equation~\ref{EQ2}) based on  {\sl Gaia} EDR3 magnitudes, 
together with corresponding reference mean colour used for the fit. 
 The uncertainties for each coefficients are listed in brackets.}             % title of Table
\label{tab1alt}      % is used to refer this table in the text
\centering                          % used for centering table
\begin{tabular}{c c     c c c  c c c}        % centered columns (4 columns)
\hline\hline                 % inserts double horizontal lines
{\rm Colour} & {\rm $<$Colour$>$} &   
    ${\rm c_0}$ & ${\rm c_1}$ & ${\rm c_2}$ & ${\rm c_3}$ & 
    ${\rm c_4}$ & ${\rm c_5}$  \\ 
\hline
	  &         &    &   &      &     &      &    \\
\hline
{\bf Dwarf stars}	      &         &  &   &      &     &      &       \\
\hline
\bprp\  &  0.8    & 0.8918   &  0.5120   &  -0.0353    &   -0.0065      &  -0.0020    &	 -0.0395      	   \\
        &         &(0.0007) & (0.0041) & (0.0081) & (0.0007) & (0.0006) & (0.0032)\\ 
\bpg\   &  0.3    & 0.8798   &  1.0774   &  -0.4677    &   -0.0065      &  -0.0031	  &  -0.0752 	    \\
        &         &(0.0007) & (0.0084) & (0.0305) & (0.0007) & (0.0006) & (0.0064)\\ 
\grp\   &  0.5    & 0.9016   &  0.9905   &   0.2284    &   -0.0069      &  -0.0011    &  -0.0726 	    \\
        &         &(0.0008) & (0.0079) & (0.0330) & (0.0008) & (0.0007) & (0.0064)\\ 
\bpk\   &  1.8    & 0.8977   &  0.2204   &  -0.0021    &   -0.0024      &	0.0005    &  -0.0158  	    \\
        &         &(0.0005) & (0.0013) & (0.0011) & (0.0006) & (0.0005) & (0.0010)\\ 
\rpk\   &  0.9    & 0.8636   &  0.3824   &  -0.0121    &    0.0045      &	0.0029    &  -0.0220 	    \\
        &         &(0.0007) & (0.0030) & (0.0044) & (0.0006) & (0.0006) & (0.0021)\\ 
\gk\    &  1.5    & 0.9015   &  0.2733   &   0.0016    &    0.0000      &	0.0015    &  -0.0163 	    \\
        &         &(0.0006) & (0.0018) & (0.0020) & (0.0006) & (0.0005) & (0.0014)\\ 
\hline
{\bf Giant stars}	     &      &   &   &   &      &     &             \\
\hline
\bprp\  &  1.1    &  1.0294   &  0.4031    &     -0.0344    &  -0.0059    &	 -0.0020   &   	-0.0009    \\
        &         &(0.0020) & (0.0059) & (0.0136) & (0.0012) & (0.0011) & (0.0050)\\ 
\bpg\   &  0.5    &  1.0582   &  0.7372    & -0.3710    &     -0.0085    &  -0.0039	    &  0.0070 	    \\
        &         &(0.0021) & (0.0115) & (0.0507) & (0.0012) & (0.0011) & (0.0096)\\ 
\grp\   &  0.6    &  0.9962   &  0.8640    &  0.2347          &  -0.0046    &	-0.0012    &   0.0060	    \\
        &         &(0.0018) & (0.0107) & (0.0511) & (0.0011) & (0.0009) & (0.0091)\\ 
\bpk\   &  2.5    &  1.0338   &  0.1872    &  -0.0017       &   -0.0023   &	 -0.0008   &   -0.0045	    \\
        &         &(0.0012) & (0.0017) & (0.0015) & (0.0007) & (0.0006) & (0.0014)\\ 
\rpk\   &  1.4    &  1.0382   &  0.3334    &  -0.0226         &  0.0016    &  0.0007	    &  -0.0221    \\
        &         &(0.0015) & (0.0038) & (0.0061) & (0.0008) & (0.0008) & (0.0032)\\ 
\gk\    &  2.0    &  1.0265   &  0.2429    &  -0.0035       &  0.0014    &	 0.0007   &   -0.0089	    \\
        &         &(0.0011) & (0.0021) & (0.0025) & (0.0006) & (0.0006) & (0.0018)\\ 
\hline                  
\hline                                   %inserts single line
\end{tabular}
\end{table*}

\begin {itemize}
\item {\sl \bprp\ - dwarf stars}
\tiny 
\[
\left[
\begin{array}{rrrrrr}
 1.000 &  0.045 & -0.229 & -0.034 & -0.503 &  0.075\\ 
 0.008 &  1.000 & -0.288 & -0.414 &  0.359 & -0.699\\ 
-0.021 & -0.288 &  1.000 & -0.012 & -0.114 & -0.198\\ 
-0.035 & -0.414 & -0.012 &  1.000 & -0.615 &  0.350\\ 
-0.583 &  0.359 & -0.114 & -0.615 &  1.000 & -0.400\\ 
 0.017 & -0.699 & -0.198 &  0.350 & -0.400 &  1.000\\ 
\end{array}
\right]
\]
\normalsize
\item {\sl \bprp\ - giant stars}
\tiny
\[
\left[
\begin{array}{rrrrrr}
 1.000 & -0.070 & -0.335 &  0.001 & -0.651 & -0.013\\ 
-0.024 &  1.000 &  0.110 & -0.281 &  0.056 & -0.387\\ 
-0.050 &  0.110 &  1.000 & -0.342 & -0.124 &  0.108\\ 
 0.003 & -0.281 & -0.342 &  1.000 &  0.281 & -0.050\\ 
-1.252 &  0.056 & -0.124 &  0.281 &  1.000 & -0.279\\ 
-0.005 & -0.387 &  0.108 & -0.050 & -0.279 &  1.000\\ 
\end{array}
\right]
\]
\normalsize
\item {\sl \bpg\ - dwarf stars}
\tiny
\[
\left[
\begin{array}{rrrrrr}
 1.000 & -0.079 & -0.175 &  0.018 & -0.578 &  0.192\\ 
-0.007 &  1.000 & -0.416 & -0.330 &  0.331 & -0.597\\ 
-0.004 & -0.416 &  1.000 &  0.032 & -0.061 & -0.272\\ 
 0.020 & -0.330 &  0.032 &  1.000 & -0.580 &  0.215\\ 
-0.679 &  0.331 & -0.061 & -0.580 &  1.000 & -0.401\\ 
 0.022 & -0.597 & -0.272 &  0.215 & -0.401 &  1.000\\ 
\end{array}
\right]
\]

\normalsize
\item {\sl \bpg\ - giant stars}
\tiny
\[
\left[
\begin{array}{rrrrrr}
 1.000 &  0.068 & -0.340 & -0.067 & -0.638 & -0.046\\ 
 0.012 &  1.000 &  0.222 & -0.432 &  0.039 & -0.391\\ 
-0.014 &  0.222 &  1.000 & -0.293 & -0.081 & -0.018\\ 
-0.115 & -0.432 & -0.293 &  1.000 &  0.193 &  0.223\\ 
-1.238 &  0.039 & -0.081 &  0.193 &  1.000 & -0.301\\ 
-0.010 & -0.391 & -0.018 &  0.223 & -0.301 &  1.000\\

\end{array}
\right]
\]
\normalsize
\item {\sl \grp\ - dwarf stars}
\tiny
\[
\left[
\begin{array}{rrrrrr}
 1.000 &  0.137 & -0.268 & -0.085 & -0.435 & -0.052\\ 
 0.014 &  1.000 & -0.136 & -0.498 &  0.368 & -0.774\\ 
-0.006 & -0.136 &  1.000 & -0.033 & -0.144 & -0.102\\ 
-0.084 & -0.498 & -0.033 &  1.000 & -0.641 &  0.459\\ 
-0.514 &  0.368 & -0.144 & -0.641 &  1.000 & -0.404\\ 
-0.006 & -0.774 & -0.102 &  0.459 & -0.404 &  1.000\\

\end{array}
\right]
\]
\normalsize
\item {\sl \grp\ - giant stars}
\tiny
\[
\left[
\begin{array}{rrrrrr}
 1.000 & -0.226 & -0.345 &  0.060 & -0.630 &  0.019\\ 
-0.038 &  1.000 & -0.001 & -0.074 &  0.077 & -0.400\\ 
-0.012 & -0.001 &  1.000 & -0.434 & -0.161 &  0.238\\ 
 0.100 & -0.074 & -0.434 &  1.000 &  0.354 & -0.349\\ 
-1.263 &  0.077 & -0.161 &  0.354 &  1.000 & -0.256\\ 
 0.004 & -0.400 &  0.238 & -0.349 & -0.256 &  1.000\\

\end{array}
\right]
\]

\normalsize
\item {\sl \bpk\ - dwarf stars}
\tiny
\[
\left[
\begin{array}{rrrrrr}

 1.000 &  0.098 & -0.243 & -0.073 & -0.425 & -0.035\\ 
 0.041 &  1.000 & -0.205 & -0.451 &  0.386 & -0.805\\ 
-0.118 & -0.205 &  1.000 & -0.013 & -0.181 & -0.026\\ 
-0.068 & -0.451 & -0.013 &  1.000 & -0.678 &  0.435\\ 
-0.471 &  0.386 & -0.181 & -0.678 &  1.000 & -0.393\\ 
-0.018 & -0.805 & -0.026 &  0.435 & -0.393 &  1.000\\ 

\end{array}
\right]
\]

\normalsize
\item {\sl \bpk\ - giant stars}
\tiny
\[
\left[
\begin{array}{rrrrrr}
 1.000 & -0.021 & -0.313 & -0.038 & -0.680 & -0.064\\ 
-0.015 &  1.000 &  0.135 & -0.278 &  0.037 & -0.499\\ 
-0.246 &  0.135 &  1.000 & -0.358 & -0.111 &  0.133\\ 
-0.067 & -0.278 & -0.358 &  1.000 &  0.262 & -0.001\\ 
-1.327 &  0.037 & -0.111 &  0.262 &  1.000 & -0.206\\ 
-0.055 & -0.499 &  0.133 & -0.001 & -0.206 &  1.000\\

\end{array}
\right]
\]

\normalsize
\item {\sl \rpk\ - dwarf stars}
\tiny
\[
\left[
\begin{array}{rrrrrr}
 1.000 & -0.182 & -0.122 &  0.012 & -0.531 &  0.197\\ 
-0.040 &  1.000 & -0.434 & -0.222 &  0.382 & -0.761\\ 
-0.018 & -0.434 &  1.000 & -0.003 & -0.228 &  0.035\\ 
 0.012 & -0.222 & -0.003 &  1.000 & -0.648 &  0.179\\ 
-0.601 &  0.382 & -0.228 & -0.648 &  1.000 & -0.335\\ 
 0.061 & -0.761 &  0.035 &  0.179 & -0.335 &  1.000\\ 
\end{array}
\right]
\]

\normalsize
\item {\sl \rpk\ - giant stars}
\tiny
\[
\left[
\begin{array}{rrrrrr}
 1.000 & -0.021 & -0.315 & -0.033 & -0.688 & -0.081\\ 
-0.008 &  1.000 &  0.176 & -0.271 &  0.054 & -0.512\\ 
-0.077 &  0.176 &  1.000 & -0.346 & -0.094 &  0.169\\ 
-0.060 & -0.271 & -0.346 &  1.000 &  0.226 &  0.054\\ 
-1.353 &  0.054 & -0.094 &  0.226 &  1.000 & -0.176\\ 
-0.038 & -0.512 &  0.169 &  0.054 & -0.176 &  1.000\\

\end{array}
\right]
\]

\normalsize
\item {\sl \gk\ - dwarf stars}
\tiny
\[
\left[
\begin{array}{rrrrrr}
 1.000 &  0.166 & -0.225 & -0.111 & -0.407 & -0.116\\ 
 0.056 &  1.000 & -0.102 & -0.457 &  0.328 & -0.823\\ 
-0.068 & -0.102 &  1.000 & -0.028 & -0.206 & -0.018\\ 
-0.105 & -0.457 & -0.028 &  1.000 & -0.663 &  0.448\\ 
-0.468 &  0.328 & -0.206 & -0.663 &  1.000 & -0.341\\ 
-0.050 & -0.823 & -0.018 &  0.448 & -0.341 &  1.000\\ 

\end{array}
\right]
\]

\normalsize
\item {\sl \gk\ - giant stars}
\tiny
\[
\left[
\begin{array}{rrrrrr}
 1.000 & -0.057 & -0.323 & -0.013 & -0.680 & -0.072\\ 
-0.031 &  1.000 &  0.136 & -0.220 &  0.035 & -0.502\\ 
-0.147 &  0.136 &  1.000 & -0.383 & -0.114 &  0.200\\ 
-0.023 & -0.220 & -0.383 &  1.000 &  0.277 & -0.070\\ 
-1.344 &  0.035 & -0.114 &  0.277 &  1.000 & -0.175\\ 
-0.046 & -0.502 &  0.200 & -0.070 & -0.175 &  1.000\\ 

\end{array}
\right]
\]
\end{itemize}

\end{appendix}
\end{document}